# The Occurrence and Mass Distribution of Close-in Super-Earths, Neptunes, and Jupiters

Andrew W. Howard,[1,2]* Geoffrey W. Marcy,[1] John Asher Johnson,[3]
Debra A. Fischer,[4] Jason T. Wright,[5] Howard Isaacson,[1]
Jeff A. Valenti,[6] Jay Anderson,[6] Doug N. C. Lin,[7,8] Shigeru Ida[9]

[1]Department of Astronomy, University of California, Berkeley, CA 94720, USA

[2]Townes Fellow, Space Sciences Laboratory, University of California, Berkeley, CA 94720, USA

[3]Department of Astrophysics, California Institute of Technology, Pasadena, CA 91125, USA

[4]Department of Astronomy, Yale University, New Haven, CT 06511, USA

[5]Department of Astronomy & Astrophysics, The Pennsylvania State University,
University Park, PA 16802, USA

[6]Space Telescope Science Institute, 3700 San Martin Dr., Baltimore, MD 21218, USA

[7]UCO/Lick Observatory, University of California, Santa Cruz, CA 95064, USA

[8]Kavli Institute for Astronomy and Astrophysics, Peking University, Beijing, China

[9]Tokyo Institute of Technology, Ookayama, Meguro-ku, Tokyo 152-8551, Japan

*To whom correspondence should be addressed; E-mail: howard@astro.berkeley.edu.

**The questions of how planets form and how common Earth-like planets are can be addressed by measuring the distribution of exoplanet masses and orbital periods. We report the occurrence rate of close-in planets (with orbital periods less than 50 days) based on precise Doppler measurements of 166 Sun-like stars. We measured increasing planet occurrence with decreasing planet mass ($M$). Extrapolation of a power law mass distribution fitted to our measurements, $df/d\log M = 0.39 M^{-0.48}$, predicts that 23% of stars harbor a close-in Earth-mass planet (ranging from 0.5 to 2.0 Earth-masses). Theoretical models of planet formation predict a deficit of planets in the domain from 5 to 30 Earth-masses and with orbital periods less than 50 days. This region of parameter space is in fact well populated, implying that such models need substantial revision.**

The architecture of our Solar System, with small rocky planets orbiting close to the Sun and gas-liquid giant planets farther out, provides key properties that inform theories of planet formation and evolution. As more planetary systems are discovered the planet occurrence fractions and distributions in mass and orbital distance similarly shape our understanding of how planets form, interact, and evolve. Such properties can be measured using precise Doppler measurements of the host stars that interact gravitationally with their planets. These measurements reveal the planetary orbits and minimum masses ($M \sin i$, due to unknown orbital inclinations $i$).

In the core-accretion theory of planet formation, planets are built from the collisions and sticking together of rock-ice planetesimals, growing to Earth-size and beyond, followed by gravitational accretion of hydrogen and helium gas. This process has been simulated numerically (1–4), predicting the occurrence of planets in a two-parameter space defined by their masses and orbital periods. These simulations predict that there should be a paucity of planets, a "planet desert" (*3*), in the mass range ~1–30 Earth-masses ($M_{Earth}$) orbiting inside of ~1 astronomical unit (AU), depending on the exact treatment of inward planet migration.

We use precise Doppler measurements of a well-defined sample of nearby stars to detect planets having masses of 3–1000 $M_{Earth}$ orbiting within the inner 0.25 AU. The 235 main sequence G, K, and M-type dwarfs stars in our NASA-UC Eta-Earth Survey were selected from the Hipparcos catalog based on brightness ($V < 11$), distance ($< 25$ pc), luminosity ($M_V > 3.0$), low chromospheric activity ($\log R'_{HK} < -4.7$), lack of stellar companions, and observability from Keck Observatory. The resulting set of stars is nearly free of selection bias; in particular, stars were neither included nor excluded based on their likelihood to harbor a planet. (The stars and planets are listed in the Supporting Online Material, hereafter SOM.) Here we focus on the 166 G and K-type stars, with masses of 0.54–1.28 solar masses and $B - V < 1.4$. We analyze previously announced planets, new candidate planets, and non-detections on a star-by-star basis to measure close-in planet occurrence as a function of planet mass.

We measured at least 20 radial velocities (RV) for each star, achieving 1 m s$^{-1}$ precision (*5*) with the HIRES echelle spectrometer (*6*) at Keck Observatory. To achieve sensitivity on timescales from years to days, the observations of each star were spread over 5 years, with at least one cluster of 6–12 observations in a 12 night span. Stars with candidate planets were observed intensively leading to several discoveries (*5*, *7*, *8*). In total, 33 planets (Fig. 1) have been detected around 22 stars in our sample (*5*, *7*, *9–23*), some of which were discovered by other groups. Sixteen of these planets have $P < 50$ days. Our analysis also includes five candidate low-mass planets from the Eta-Earth Survey with $P < 50$ days and false alarm probabilities (*5*) of $< 5\%$.

For each star without a detected or candidate planet, we computed on a fine grid of orbital periods the maximum mass planet compatible with the RV measurements. We assumed circu-lar orbits and removed linear trends and stellar activity correlations (when appropriate) before fitting. Above this mass limit at a particular period, a planet would be detected; below it, the existence of a planet cannot be ruled out. Folding together these limits for all stars, we derived a search completeness function, $C(P, M\sin i)$. As shown in Fig. 1, $C$ is the fraction of stars with sufficient measurements to rule out planets of a given minimum mass and orbital period.

We computed the occurrence rate—the fraction of stars orbited by planets—in five planet-mass domains restricted to orbital periods of $P < 50$ days (see Fig. 2). In the three largest mass domains our survey is complete because these planets impart easily detectable Doppler signatures ($K > 9$ m s$^{-1}$). In the two lowest mass domains, there are markedly higher planet occurrence rates, despite reduced sensitivity (depicted by the shaded regions in Fig. 1). We corrected for this incompleteness by computing a "missing planet correction" by sampling $C(P, M\sin i)$ for each detected and candidate planet. The fraction of stars capable of revealing each planet is $C$; therefore, for each detected or candidate planet, $1/C - 1$ missed planets remain undetected. Summing over detected and candidate planets, we estimate that 10.2 and 4.6 planets were missed in the 3–10 and 10–30 $M_{\text{Earth}}$ mass domains, respectively. Including these missed planets and using binomial statistics (see SOM), we computed the planet occurrence rates shown in Fig. 2. There is a substantial increase in planet occurrence with decreasing planet mass.

We fit the planet occurrence rate in five mass domains to a power law, $\mathrm{d}f/\mathrm{d}\log M_E = kM_E^\alpha$, where $\mathrm{d}f/\mathrm{d}\log M_E$ is the occurrence rate in a $\log_{10}$ mass interval, $M_E = M\sin i/M_{\text{Earth}}$, $k = 0.39^{+0.27}_{-0.16}$, and $\alpha = -0.48^{+0.12}_{-0.14}$, Using this model, we extrapolated speculatively into two important mass domains below our detection limit. We expect planets of mass 1–3 $M_{\text{Earth}}$ to orbit $14^{+8}_{-5}$% of Sun-like stars. For Earth-like planets with $M\sin i = 0.5$–2 $M_{\text{Earth}}$ and $P < 50$ days, we predict an occurrence rate of $\eta_{\text{Earth}} = 23^{+16}_{-10}$%.

Our measurements in the two largest mass domains are consistent with a 1.2%±0.2% occurrence rate of "hot Jupiters" within 0.1 AU ($P < \sim12$ d) for FGK dwarfs (24). At lower masses, our results are consistent with the two planets in the mass range $M \sin i$ = 5–30 $M_{Earth}$ with $P < 16$ days detected around 24 FGK dwarfs surveyed by the Anglo-Australian Telescope (25). Mayor et al. have noted a substantially higher planet occurrence rate, from 30% ± 10% (26) to "at least 50%" (27), for planets with $M \sin i$ = 3–30 $M_{Earth}$ and $P < 50$ days around GK stars based on measurements with the HARPS spectrometer. Accounting for missed planets, we find an occurrence rate of $15^{+5}_{-4}$% with a 24% upper limit (95% confidence) for this range of parameters.

Our analysis extends to lower masses the work of Cumming et al. (28) who measured 10.5% of Solar-type stars hosting a gas-giant planet ($M \sin i$ = 100–3000 $M_{Earth}$, $P$ = 2–2000 d) with planet occurrence varying as $df \propto M^{-0.31\pm0.2}P^{0.26\pm0.1}\,d\log M\,d\log P$. Although the details of planet formation probably differ for gas-giant and terrestrial planets, we can speculate that if the trend of increasing planet occurrence with longer orbital period holds down to an Earth mass, then $\eta_{Earth} = 23$% is an underestimate for orbits near 1 AU. For orbits beyond the ice line (~2.5 AU), gravitational microlensing searches find three times as many Neptunes as Jupiters (29), suggesting that planet occurrence also increases with decreasing planet mass in this domain.

The distribution of planets in the mass/orbital-period plane (Fig. 1) reveals important clues about planet formation and migration. Planets with $M \sin i$ = 10–100 $M_{Earth}$ and $P < 20$ days are almost entirely absent. There is also an over-density of planets starting at $P < 10$ days and $M \sin i$ = 4–10 $M_{Earth}$ and extending to higher masses and longer periods. These patterns suggest different formation and migration mechanisms for close-in low-mass planets compared to massive gas-giant planets.

Population synthesis models of planet formation predict an increase in planet occurrence with decreasing planet mass (2, 30). However, the bulk of their predicted low-mass planets reside near the ice line, well outside of the $P < 50$ days domains analyzed here. In fact, these models predict a "planet desert" precisely in the domain of mass and period where we detect an over-density of planets. The desert emerges naturally in the simulations (*3, 4*) from fast migration and accelerating planet growth. Most planets are born near or beyond the ice line and those that grow to a critical mass of several Earth-masses either rapidly spiral inward to the host star or undergo runaway gas accretion and become massive gas-giants. Our measurements show that population synthesis models of planet formation are currently inadequate to explain the distribution of low-mass planets.

The Kepler mission (*31*) is currently surveying 156,000 faint stars for transiting planets as small as the Earth. Our power law model predicts that Kepler will detect a bounty of close-in small planets: an occurrence rate of 22% for $P < 50$ days and $M \sin i$ = 1–8 $M_{\text{Earth}}$, corresponding to 1–2 Earth-radii assuming terrestrial, Earth-like density (5500 kg m$^{-3}$). When the mission is complete, we estimate (see SOM) that Kepler will have detected the transits of 120–260 of these plausibly terrestrial worlds orbiting the $\sim 10^4$ G and K dwarfs brighter than 13th magnitude (*32, 33*).

34. This work was based on observations at the W. M. Keck Observatory granted by NASA and the University of California. We thank the many observers who contributed to the measurements reported here, and acknowledge the efforts and dedication of the Keck Observatory staff, especially S. Dahm, H. Tran, and G. Hill for support of HIRES and G. Wirth for support of remote observing. We acknowledge R. P. Butler and S. Vogt for many years of contributing to the data presented here. A. H. acknowledges support from a Townes Post-doctoral Fellowship at the U. C. Berkeley Space Sciences Laboratory. G. M. acknowledges NASA grant NNX06AH52G. Finally, we extend special thanks to those of Hawaiʻian ancestry on whose sacred mountain of Mauna Kea we are privileged to be guests.

Supporting Online Material

www.sciencemag.org

Materials and Methods

Figs. S1, S2, S3, S4, S5, S6, S7

Tables S1, S2, S3

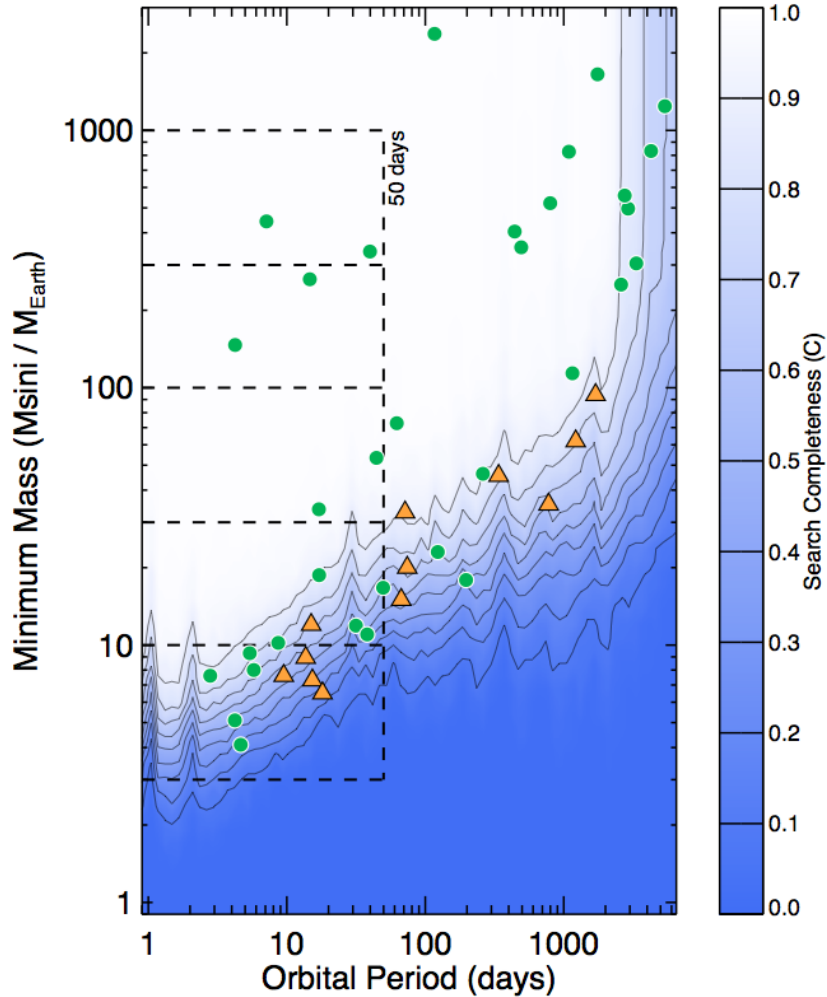

Figure 1: Detected planets (green circles) and candidate planets (orange triangles) from the Eta-Earth Survey as a function of orbital period and minimum mass. Five mass domains—3–10, 10–30, 30–100, 100–300, 300-1000 Earth-masses—out to 50 day orbits are marked with dashed lines. Search completeness—the fraction of stars with sufficient measurements to rule out planets in circular orbits of a given minimum mass and orbital period—is shown as blue contours from 0.0 to 1.0 in steps of 0.1.

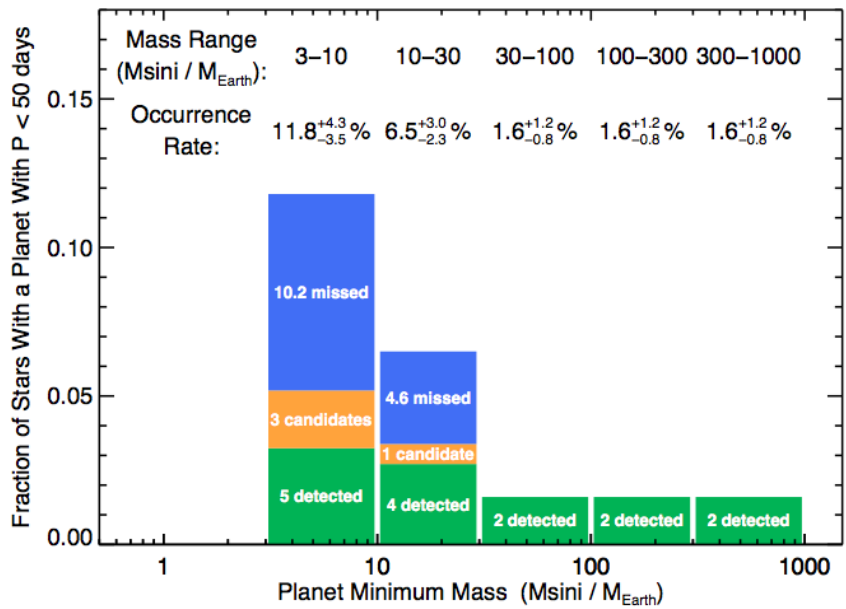

Figure 2: Histogram of the occurrence rate of stars hosting planets with orbital periods of less than 50 days in five mass ranges. Detected (green), candidate (orange), and missed (blue) planets are depicted separately. Missed planets statistically correct for planets that are detectable by $1~\mathrm{m\,s^{-1}}$ measurements, but were missed because of non-uniform sensitivity.

# Materials and Methods

In this supplement to "The Occurrence and Mass Distribution of Close-in Super-Earths, Neptunes, and Jupiters" by Howard et al., we provide materials and methods for our analysis. We describe the radial velocity (RV) measurements at Keck observatory (Section S1) and the Eta-Earth Survey stars and planets (Section S2). We discuss calculations to compute limits on the mass of planetary companions for each star and how these limits are folded together to compute a search completeness function, $C$ (Section S3). Our methods for computing occurrence rates as a function of planet mass are described in Section S4. In Section S5, we calculate the number of stars detectable by the Kepler mission with orbital period $P < 50$ d and mass 1–8 $M_{\rm Earth}$.

## S1  Observations and Doppler Analysis

We observed the Eta-Earth Survey stars in Table S1 using the HIRES echelle spectrometer (*1*) on the 10-m Keck I telescope. Exposure times ranged from a few seconds for the brightest stars to 500 seconds for the faintest. All observations were made with an iodine cell mounted directly in front of the spectrometer entrance slit. The dense set of molecular absorption lines imprinted on the stellar spectra provide a robust wavelength fiducial against which Doppler shifts are measured, as well as strong constraints on the shape of the spectrometer instrumental profile at the time of each observation (*2, 3*).

We measured the Doppler shift from each star-times-iodine spectrum using a forward modeling procedure modified from the method described in (*4*). The most significant modification is the way we model the intrinsic stellar spectrum, which serves as a reference point for the relative Doppler shift measurements for each observation. Butler et al. use a version of the Jansson deconvolution algorithm (*5*) to remove the spectrometer's instrumental profile from an iodine-free template spectrum. We instead use a deconvolution algorithm that employs a more effective regularization scheme, which results in significantly less noise amplification and improved Doppler precision.

All of the measurements for this analysis were made after 2004 Aug. when HIRES was upgraded with a new CCD and started to consistently achieve 1 m s$^{-1}$ precision (*6*). Many of the Eta-Earth Survey stars also have "pre-upgrade" measurements of lower quality that are not used here.

For each observation we also measure $S_{\rm HK}$ and the related quantity $\log R'_{\rm HK}$ from the flux in the cores of the Ca II H and K lines in the HIRES spectra (*7*). These indicators of chromospheric activity provide useful checks that detected RV variations are due to a planet in orbital motion and not magnetic activity on the stellar surface modulated by stellar rotation and activity cycles. We identified 13 stars with significant correlations between RV and $S_{\rm HK}$. For these stars, we de-correlated the activity signal by subtracting from the RVs a linear fit to the measured RVs as a function of $S_{\mathrm{HK}}$. Typically this process removes 2–4 m s$^{-1}$ of RV variability.

## S2  Eta-Earth Survey Target Stars and Planets

We selected 235 G, K, and M-type dwarf stars for the NASA-UC Eta-Earth Survey. This survey was designed as a search for planets orbiting a well-controlled sample of stars with the goal of characterizing the population of planets using detections and non-detections. The stars were selected from the Hipparcos catalog (8) based on brightness ($V < 11$), distance ($< 25$ pc), luminosity ($M_V > 3.0$), chromospheric activity ($\log R'_{\mathrm{HK}} < -4.7$), lack of stellar companions, and observability from Keck Observatory. The resulting set of stars is nearly free of selection bias; in particular, stars were neither included nor excluded based on their likelihood to harbor a planet. Metallicity was not used as a selection criterion. Here we focus on the 166 G and K-type stars with masses in the range 0.54–1.28 solar masses and $B - V < 1.4$. Table S1 lists the names of these stars (using HD or Hipparcos identifiers), their spectral types, masses, and the number of observations for each star. Figure S1 shows the distribution of stellar masses.

Each star was observed over 5 years and during at least one "high-cadence run" of 6–12 observations in a 12 night span to increase our sensitivity to short-period signals. For some stars with candidate planets, we made 2–5 consecutive observations and averaged the resulting velocities for a single measurement. Figure S2 shows a histogram of the number of RV measurements per star. We made at least 20 RV (and as many as 144) measurements for each star; the median number of measurements is 33. Stars with substantially more than the median number of observations typically have detected planets or particularly low astrophysical jitter. Additionally, when a candidate planet with a formal false alarm probability (FAP) (6) of $< 20\%$ was identified, the host star was observed intensively as telescope time permitted.

Table S2 lists the 33 planets detected around 22 Eta-Earth Survey stars, their orbital periods, minimum masses, and references in the literature. We use HD names to identify the stars hosting each planet to match the star names in Table S1. Some of these planets were initially detected by other groups, but were confirmed by our HIRES measurements. We do not announce any new planets here. Table S3 lists the orbital periods and minimum masses ($M \sin i$) of the 12 candidate planets orbiting Eta-Earth Survey stars. The candidates have formal FAPs of $< 5\%$. We typically publish planets when their FAPs drop below 1% and we are sure that the periodic signal is due to a planet and not astrophysical noise or systematic errors.

The 33 detected planets and 12 candidate planets (45 planets in total) orbit 32 unique stars. Eight of the 32 stars have multiple detected planets (25% multiplicity). Of the 16 stars with a detected planet of any mass in a $P < 50$ d orbit, seven have multiple detected planets (44% multiplicity).

# S3  Search Completeness

Every star in the Eta-Earth Survey has a different observing history. Figure S3 shows the RV time series for a star with a large number of RV measurements (HD 185144, top) and for a star with a more sparse set of observations (HD 84737, bottom).

Figure S4 shows the distribution of velocity RMS for Eta-Earth Survey stars, excluding stars with detected planets (Table S2). Long-term RV trends were fit for and subtracted before computing the velocity RMS. Activity correlations were also subtracted for 13 stars before computing the RMS.

For each star we estimate the maximum $M \sin i$ value compatible with the RV measurements as a function of prospective orbital period. On a fine grid of orbital periods (2000 log-spaced trial orbital periods per $\log_{10}$ period interval) we fit the RVs to a circular orbit using a partially-linearized, least-squares fitting procedure (9) that minimized $\chi^2$. For each fit, the orbital period of the trial planet was allowed to vary within the narrow period bin. For stars with long-term RV trends, we simultaneously fit for these objects with each trial planet.

Each fit produced a best-fit Doppler semi-amplitude $K$ at the prospective orbital period $P$. We transformed $K(P)$ into $M \sin i_{\max}(P) = M_m(P)$ using masses for each star (10, 11). The black lines in Figure S5 trace $M_m(P)$ for two Eta-Earth Survey stars. For a well-observed star like HD 185144, we are sensitive to planets with masses as low as $M \sin i \approx 1\,\mathrm{M_{Earth}}$ for $P < 3$ d. For a more typical star with fewer observations (e.g., HD 84737), our measurements have sufficient precision to detect nearly all planets with $M \sin i > 10\,\mathrm{M_{Earth}}$ and $P < 50d$.

For short periods, $M_m(P)$ is a rapidly varying function of $P$ because our observations span several years, rendering nearby short periods distinguishable. This high period resolution is useful for testing for the existence of planets with specific orbital periods, but for our purpose these fine details are not helpful. So we compute $M_{m,e}(P)$, the upper envelope of $M_m(P)$ on a coarser period grid with 20 log-spaced trial orbital periods per $\log_{10}$ period interval. $M_{m,e}(P)$ is the maximum value of $M_m(P)$ within each coarse period interval. The green lines in Figure S5 show that $M_{m,e}(P)$ captures the behavior of $M_m(P)$ on a scale that is more appropriate for computing a search completeness correction.

Using $M_{m,e}(P)$ for the set of stars without detected or candidate planets we compute the search completeness function, $C(P, M \sin i)$, depicted in Figure 1 of the main paper. $C(P, M \sin i)$ measures the fraction of stars with sufficient measurements to rule out a planet of a given minimum mass and orbital period. We compute this quantity on the coarse period grid used for $M_{m,e}(P)$ and on a coarse grid of $M \sin i$ with 20 log-spaced $M \sin i$ bins per $\log_{10}$ interval. At a given $P$ and $M \sin i$, $C(P, M \sin i)$ is equal to the fraction of stars with $M_{m,e}(P) < M \sin i$. That is, $C(P, M \sin i)$ measures the fraction of stars where the limit on $M \sin i$ is below a certain value, for a given orbital period.

# S4  Planet Occurrence

To compute planet occurrence rates, we counted the number of stars hosting detected and candidate planets in each of five mass domains with $P < 50$ d: $M \sin i$ = 3–10, 10–30, 30–100, 100–300, and 300–1000 $M_{\text{Earth}}$. Multiple planets in the same mass domain that orbit the same star were only counted once, for the host star (e.g., HD 69830 is only counted once for the $M \sin i$ = 10–30 $M_{\text{Earth}}$ domain despite hosting two planets in that mass range). Planet host stars can be counted in multiple mass domains though (e.g., 55 Cnc hosts planets in the $M \sin i$ = 3–10, 3–100, and 100–300 $M_{\text{Earth}}$ domains and is counted in each of them). In addition to the five mass domains above, we also computed the planet occurrence rate in a domain with $M \sin i$ = 3–30 $M_{\text{Earth}}$ and $P < 50$ d for comparison with the estimates by Mayor et al.

Our estimates of planet occurrence depend on a "missing planet correction" computed for each detected and candidate planet. This correction is a statistical estimate of the number of planets of similar $P$ and $M \sin i$ that remain undetected in our sample. To compute this effective number of planets for each detected and candidate planet, we add $n_{\text{miss}} = 1/C(P, M \sin i) - 1$ missed planets. Since $C = 1$ over the three highest mass domains (see Figure 1), $n_{\text{miss}} = 0$ for these domains. For the lower right corner of the $M \sin i$ = 10–30 $M_{\text{Earth}}$ domain, we have $C < 1$ and compute $\Sigma n_{\text{miss}} = \Sigma(1/C(P, M \sin i) - 1) = 4.6$ missed planets by summing over detected and candidate planets in this domain. For much of the $M \sin i$ = 3–10 $M_{\text{Earth}}$ domain $C < 1$ and we compute $\Sigma n_{\text{miss}} = 10.2$ missed planets with a similar summation.

The best-fit values and error bars on planet occurrence and related parameters are computed using binomial statistics. We compute the probability distribution functions (pdfs) of the occurrence rate in each bin as the probability of drawing $n_{\text{det}} + n_{\text{cand}}$ detected and candidate planets out of $n_{\text{star}}$ stars. We multiply these occurrence rates and associated uncertainties by $(n_{\text{det}} + n_{\text{cand}} + n_{\text{miss}})/(n_{\text{det}} + n_{\text{cand}})$ to account for $n_{\text{miss}}$ missed planets. The median and 68.3% confidence intervals of these distributions are reported as the best-fit occurrence values and their "1-$\sigma$" errors.

We computed the best-fit power-law model to planet occurrence as a function of planet mass by randomly drawing occurrence rates from the pdfs of the five mass domains. For each of 30,000 trials we fit the occurrence rates to a power law model of the form $df/d \log M = kM^\alpha$. The resulting distribution of $k$ and $\alpha$ is shown in Figure S6. Each trial specifies a particular power law and predicts the occurrence rate of planets in several important mass domains. Figure S7 shows the occurrence rate pdfs in two important mass domains, 0.5–2.0 $M_{\text{Earth}}$ ("Earth-like" planets) and 1–8 $M_{\text{Earth}}$ ("Kepler planets"; see below).

Our analysis makes several assumptions worth noting. First, we restrict orbital fits to circular when computing $M_m(P)$. This reduces computational complexity and is physically well-motivated for close-in planets tidally coupled to their host stars. The detected and candidate close-in planets in the $M \sin i$ = 3–10 $M_{\text{Earth}}$ domain all have orbits that are consistent with circular or have $e < 0.2$. RV signals from slightly eccentric orbits are well approximated by circular orbits, and thus any reductions in sensitivity as measured by $M_{m,e}(P)$ and $C(P, M \sin i)$ will be small for such planets. We also note that our limits on $M \sin i$ are for a single planet. For stars with multiple undetected planets with nearly identical $K$ values, our sensitivity is reduced from the limits expressed by $M_{m,e}(P)$. Such systems are likely rare though since this scenario requires a compensation of $P$ and $M \sin i$ to produce nearly identical $K$. We searched for such systems in our existing detections and found that only three of eleven stars with a candidate or detected planet with $M \sin i$ = 3–30 $M_{\text{Earth}}$ and $P < 50$ d had a second planet in the same mass and period range with a Doppler semi-amplitude within a factor of $\sqrt{2}$ of the first planet. A further approximation is the binary nature of $M_m(P)$ and $M_{m,e}(P)$; planets are deemed detectable above this limit and undetectable below it. This sharp dividing line is averaged out though by using all 166 Eta-Earth stars to compute $C$. Our planet occurrence measurements are very weakly dependent on choices in constructing the fine and coarse period grids of $M_m(P)$ and $M_{m,e}(P)$. Finally, errors in $n_{\text{miss}}$ for a given detected or candidate planet scale as $1/C$ and can be large where the search is relatively incomplete and $C$ is small.

We considered the impact of a 10% error in $C$, resulting from one of the above approximations. For this test we used a modified search completeness $C'(P, M \sin i) = 1.1 \times C(P, M \sin i)$ and ran the statistical analysis again. (The value of $C'$ was capped at 1.0.) With this change, the number of missed planets, $n_{\text{miss}}$, in the $M \sin i$ = 3–10 and 10–30 $M_{\text{Earth}}$ bins goes from 10.2 and 4.6 planets to 8.5 and 3.6 planets, respectively. Thus a 10% change in $C$ results in 17% and 22% changes in $n_{\text{miss}}$ for the two bins. However, this effect is diluted by the detected and candidate planets in each bin, which are held fixed in this numerical experiment. The resulting effect on the planet occurrence rates is small. For $M \sin i$ = 3–10 $M_{\text{Earth}}$, the planet occurrence rate goes from $11.8^{+4.3}_{-3.5}\%$ to $10.7^{+3.9}_{-3.2}\%$ and for $M \sin i$ = 10–30 $M_{\text{Earth}}$ the occurrence rate goes from $6.5^{+3.0}_{-2.3}\%$ to $5.8^{+2.7}_{-2.1}\%$. The power law parameter change from $k = 0.39^{+0.27}_{-0.16}$, and $\alpha = -0.48^{+0.12}_{-0.14}$ to $k = 0.34^{+0.23}_{-0.14}$, and $\alpha = -0.46^{+0.13}_{-0.15}$. Thus, a 10% in the search completeness $C$ results in changes in the measured planet occurrence rates that are small compared to their statistical errors.

## S5 Kepler

It would be desirable to compare our planet occurrence measurements in the planet mass/orbital period plane with occurrence measurements by the Kepler mission (*12*) in the planet radius/orbital period plane. The current Kepler data release supports the qualitative observation that close-in small planets are much more common than large ones (*13*). Unfortunately, meaningful quantitative comparisons are not yet possible because the data are currently sequestered for 400 of the 706 stars identified as planet hosts. For the planet candidates with released data, only planet radii are available, not masses. Moreover, the initial data release only covers a 43.2 d span (robust transit detection typically requires three or more observed transit events), rendering highly incomplete planet occurrence statistics for orbital periods out to 50 d. We therefore proceed with a prediction for the number of small planets that Kepler will detect.

To estimate the number of planets detected by the Kepler mission with $P < 50$ d and radii of 1–2 Earth-radii (implying masses of 1–8 $M_{\text{Earth}}$ assuming Earth density of 5.5 kg m$^{-3}$), we must account for the period dependence of the planet occurrence rate and of the probability of a planet transiting its host star (rendering it potentially detectable by Kepler). Our power law model of planet occurrence predicts 22% occurrence for $P < 50$ d and $M \sin i = 1\text{–}8$ $M_{\text{Earth}}$. We naively assume that the distribution of these planets in orbital period is uniform in $\log P$ over $P = 1\text{–}50$ d. For a circular orbit, the probability that a planet transits depends on its orbital period and is given by $p_{\text{transit}} = R_{\text{star}}/a = (4\pi^2 R_{\text{star}}^3/GM_{\text{star}}P^2)^{1/3}$, where $M_{\text{star}}$ and $R_{\text{star}}$ are the stellar mass and radius, $a$ and $P$ are the semimajor axis and orbital period of the planet, and $G$ is the gravitational constant. We naively assign $R_{\text{star}} = R_{\text{sun}}$ for all stars in this calculation. Integrating the transit probability and planet occurrence over $P = 1\text{–}50$ d, we find that 1.2–2.6% of stars (68% confidence interval) will harbor a transiting planet with the specified parameters. We estimate that the $\sim 10^4$ G and K dwarfs brighter than 13th magnitude (*14*) are bright enough to yield a transit detection down to 1 Earth-radius. Thus we estimate that 120–260 stars host planets with radii of 1–2 Earth-radii and $P < 50$ d that are detectable by Kepler.

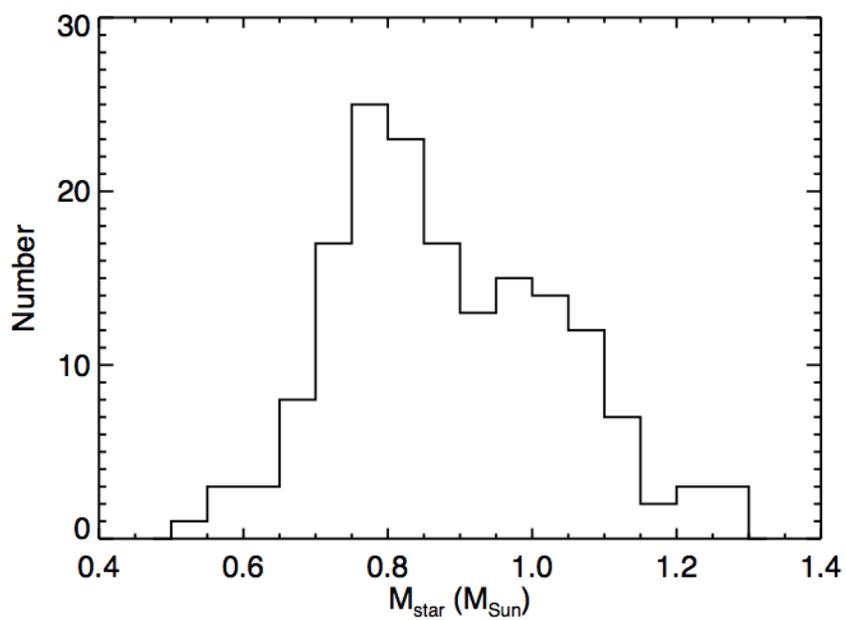

Figure S1: Histogram of stellar masses for Eta-Earth stars.

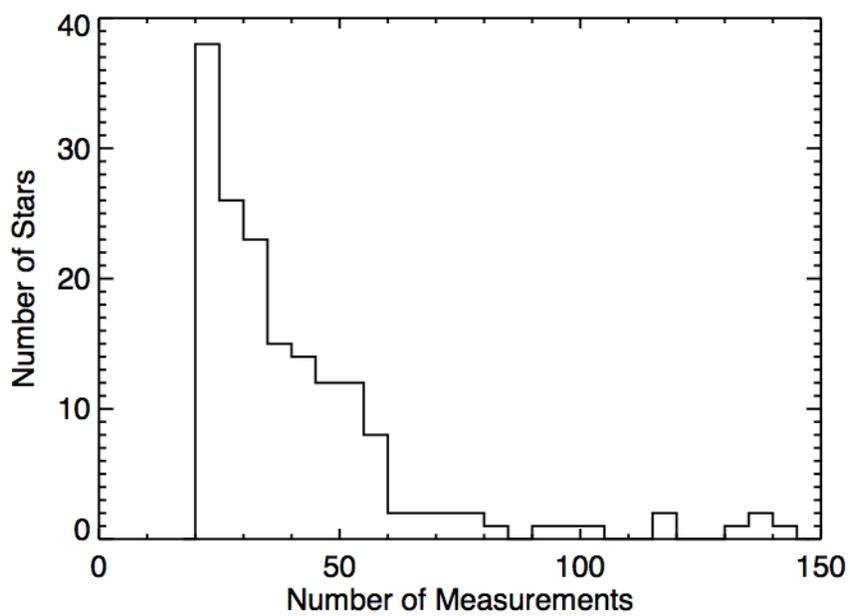

Figure S2: Histogram of number of RV measurements per Eta-Earth star.

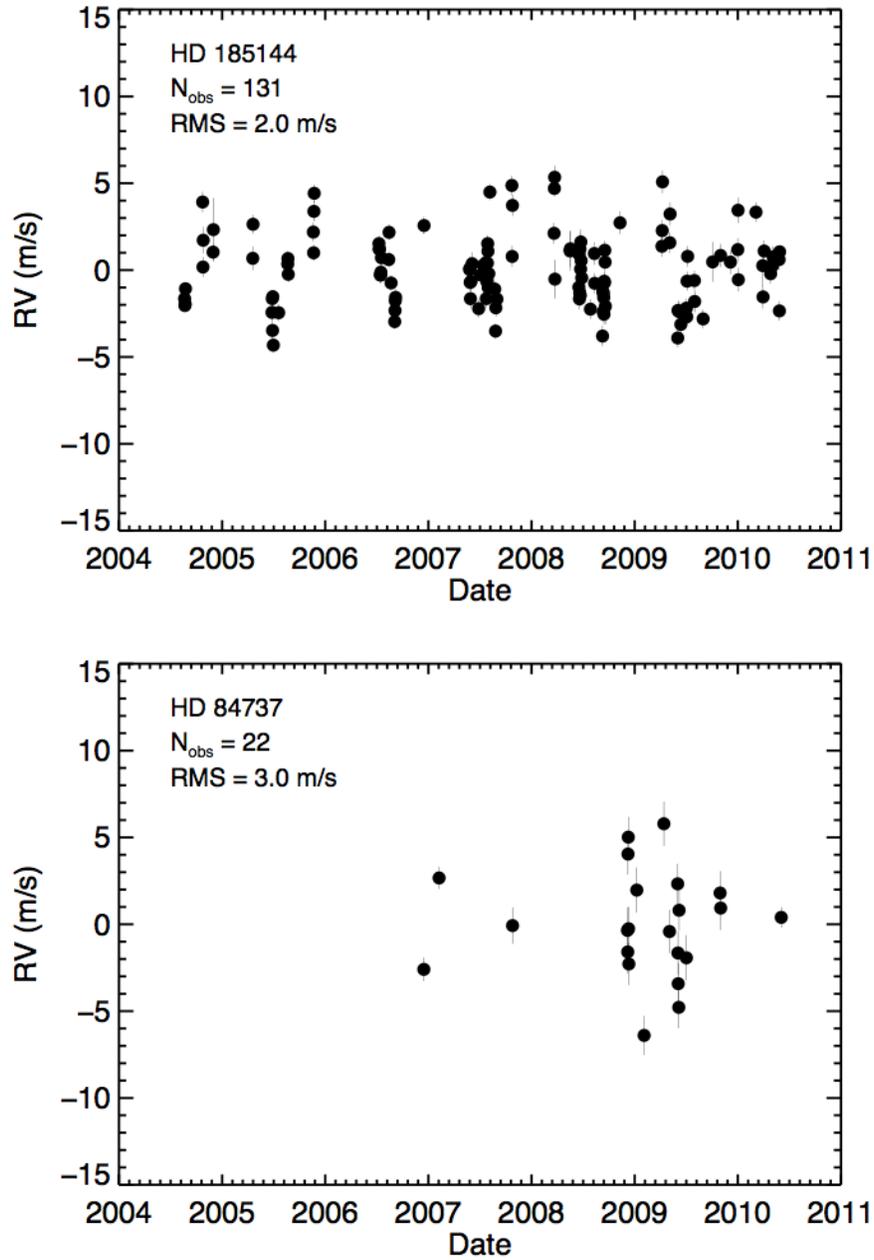

Figure S3: Radial velocity time series for a well-observed star (HD 185144, top) and a star with fewer measurements (HD 84737, bottom). The star name, number of observations, and velocity RMS are indicated in each panel.

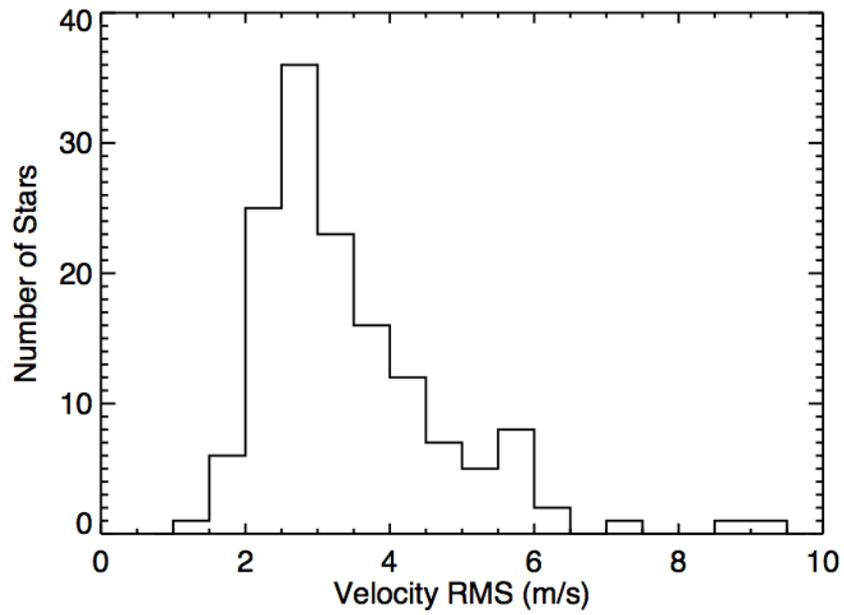

Figure S4: Histogram of the RMS of the RV measurements for each star. Stars with detected planets (Table S2) are excluded.

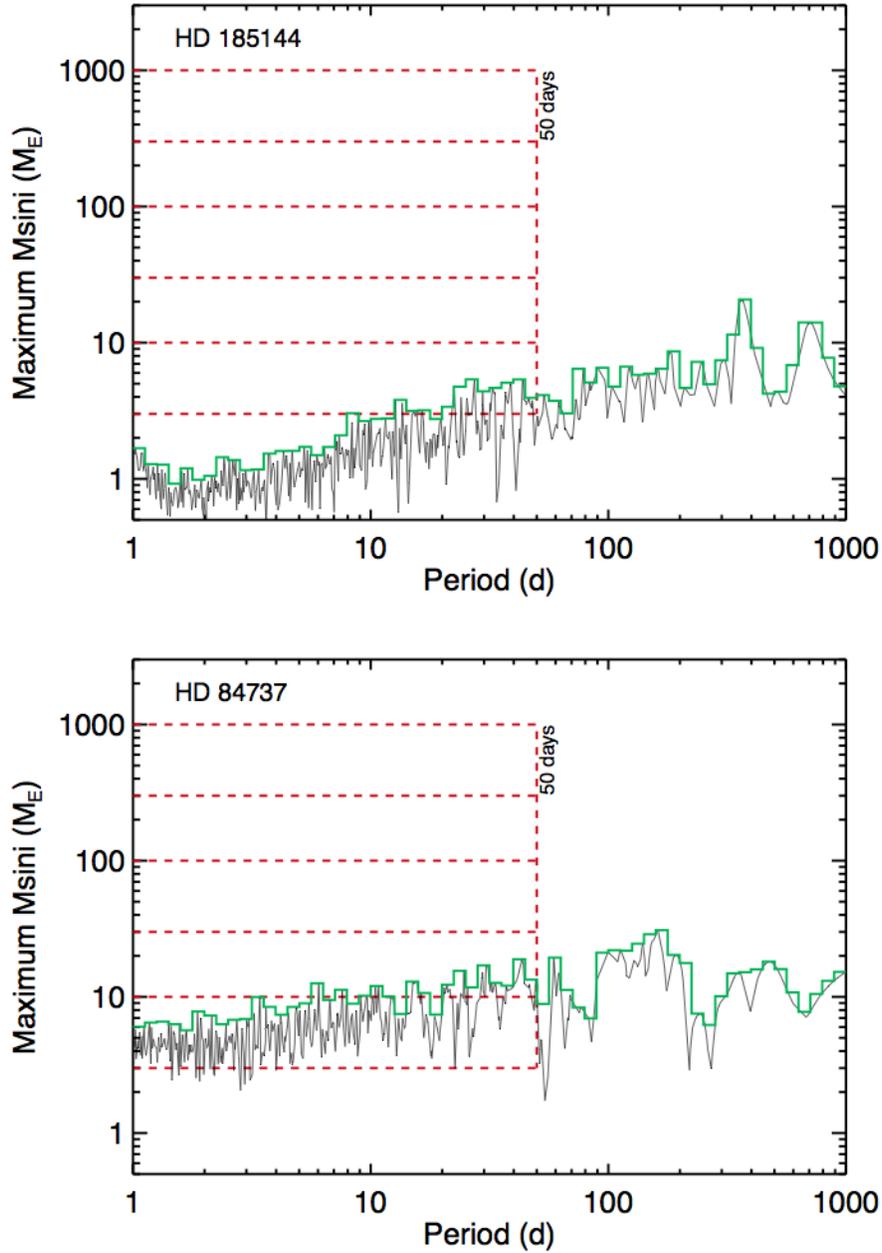

Figure S5: Limits on $M \sin i$ as a function of orbital period for the stars whose RV time series are plotted in Figure S3. The thin black lines trace this limit, $M_m(P)$, on a fine period grid. The thick green lines, $M_{m,e}(P)$, trace the upper envelope of $M_m(P)$ on a coarser period grid. Five mass domains are indicated by dashed red lines for planets with $P < 50$ d and $M \sin i = 3$–10, 10–30, 30–100, 100–300, and 300–1000 $M_{\text{Earth}}$.

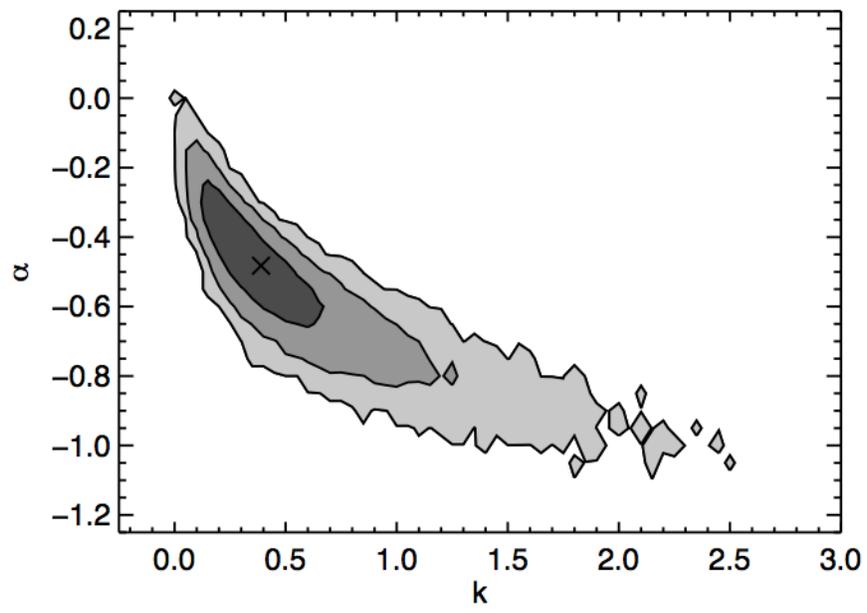

Figure S6: Probability distributions for the parameters $k$ and $\alpha$ of a power law model of planet occurrence, $\mathrm{d}f/\mathrm{d}\log M = kM^\alpha$. The X marks the best-fit values and the contours represent 68.3%, 95.4%, and 99.7% confidence intervals (1, 2, and 3-$\sigma$).

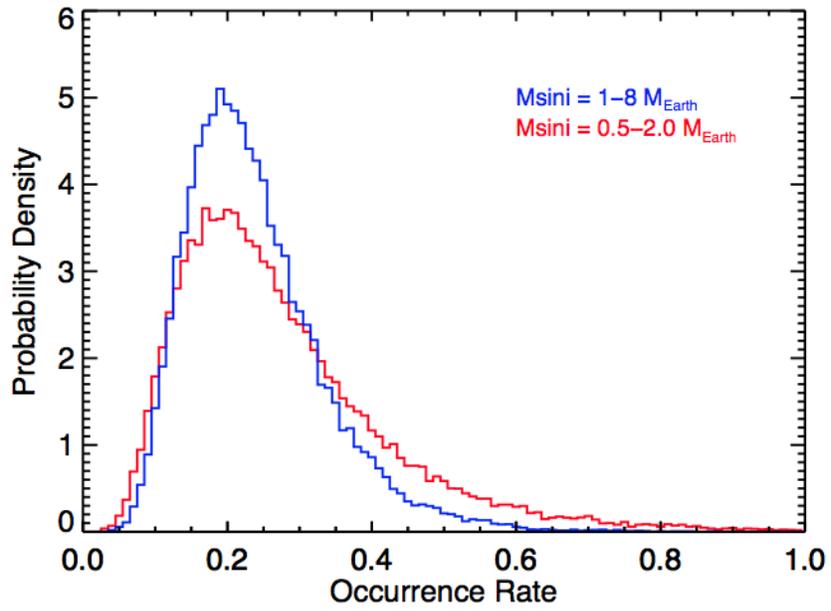

Figure S7: Probability distributions for the occurrence rate in two mass domains based on a power law model.

Table S1. G and K-type Target Stars in the Eta-Earth Survey

| Name | Spec. Type | Mass ($M_\odot$) | Num. Obs. |
|---|---|---|---|
| HD 1461 | G0 | 1.08 | 154 |
| HD 3651 | K0 | 0.89 | 29 |
| HD 3765 | K2 | 0.84 | 35 |
| HD 4256 | K2 | 0.85 | 36 |
| HD 4614 | G0 | 0.99 | 30 |
| HD 4614 B | K7 | 0.57 | 28 |
| HD 4628 | K2 | 0.72 | 49 |
| HD 4747 | G8 | 0.82 | 22 |
| HD 4915 | G0 | 0.90 | 37 |
| HD 7924 | K0 | 0.83 | 135 |
| HD 9407 | G6 | 0.98 | 97 |
| HD 10476 | K1 | 0.83 | 56 |
| HD 10700 | G8 | 0.95 | 133 |
| HD 12051 | G5 | 0.99 | 52 |
| HD 12846 | G2 | 0.88 | 36 |
| HD 14412 | G5 | 0.78 | 37 |
| HD 16160 | K3 | 0.76 | 47 |
| HD 17230 | K5 | 0.69 | 31 |
| HD 18143 | G5 | 0.90 | 35 |
| HD 18803 | G8 | 1.00 | 32 |
| HD 19373 | G0 | 1.20 | 47 |
| HD 20165 | K1 | 0.82 | 26 |
| HD 20619 | G1 | 0.91 | 35 |
| HD 22879 | F9 | 0.79 | 22 |
| HD 23356 | K2 | 0.78 | 22 |
| HD 23439 | K1 | 0.67 | 26 |
| HD 24238 | K0 | 0.73 | 29 |
| HD 24496 | G0 | 0.94 | 47 |
| HD 25329 | K1 | 0.83 | 34 |
| HD 25665 | G5 | 0.78 | 21 |
| HD 26965 | K1 | 0.78 | 41 |

Table S1—Continued

| Name | Spec. Type | Mass ($M_\odot$) | Num. Obs. |
|---|---|---|---|
| HD 29883 | K5 | 0.76 | 23 |
| HD 32147 | K3 | 0.83 | 52 |
| HD 32923 | G4 | 1.03 | 26 |
| HD 34721 | G0 | 1.12 | 21 |
| HD 34411 | G0 | 1.13 | 40 |
| HD 36003 | K5 | 0.73 | 42 |
| HD 37008 | K2 | 0.73 | 22 |
| HD 38230 | K0 | 0.83 | 24 |
| HD 38858 | G4 | 0.92 | 35 |
| HD 40397 | G0 | 0.92 | 23 |
| HD 42618 | G4 | 0.96 | 59 |
| HD 45184 | G2 | 1.04 | 46 |
| HD 48682 | G0 | 1.17 | 27 |
| HD 50692 | G0 | 1.00 | 37 |
| HD 51419 | G5 | 0.86 | 40 |
| HD 51866 | K3 | 0.78 | 32 |
| HD 52711 | G4 | 1.02 | 46 |
| HD 55575 | G0 | 1.26 | 32 |
| HD 62613 | G8 | 0.94 | 24 |
| HD 65277 | K5 | 0.72 | 21 |
| HD 65583 | G8 | 0.76 | 26 |
| HD 68017 | G4 | 0.85 | 43 |
| HD 69830 | K0 | 0.87 | 46 |
| HD 72673 | K0 | 0.78 | 23 |
| HD 73667 | K1 | 0.72 | 22 |
| HD 75732 | G8 | 0.91 | 96 |
| HD 84035 | K5 | 0.73 | 22 |
| HD 84117 | G0 | 1.15 | 22 |
| HD 84737 | G0 | 1.22 | 24 |
| HD 86728 | G3 | 1.08 | 28 |
| HD 87883 | K0 | 0.80 | 30 |

Table S1—Continued

| Name | Spec. Type | Mass ($M_\odot$) | Num. Obs. |
|---|---|---|---|
| HD 89269 | G5 | 0.89 | 29 |
| HD 90156 | G5 | 0.90 | 28 |
| HD 92719 | G2 | 1.10 | 24 |
| HD 95128 | G1 | 1.08 | 22 |
| HD 97101 | K8 | 0.60 | 21 |
| HD 97343 | G8 | 0.89 | 35 |
| HD 97658 | K1 | 0.78 | 61 |
| HD 98281 | G8 | 0.85 | 46 |
| HD 99491 | K0 | 1.01 | 71 |
| HD 99492 | K2 | 0.86 | 47 |
| HD 100180 | G0 | 1.10 | 24 |
| HD 100623 | K0 | 0.77 | 32 |
| HD 103932 | K5 | 0.76 | 44 |
| HD 104304 | G9 | 1.02 | 23 |
| HD 109358 | G0 | 1.00 | 41 |
| HD 110315 | K2 | 0.70 | 37 |
| HD 110897 | G0 | 1.23 | 29 |
| HD 114613 | G3 | 1.28 | 21 |
| HD 114783 | K0 | 0.86 | 45 |
| HD 115617 | G5 | 0.95 | 61 |
| HD 116442 | G5 | 0.76 | 25 |
| HD 116443 | G5 | 0.73 | 55 |
| HD 117176 | G4 | 1.11 | 30 |
| HD 120467 | K4 | 0.71 | 20 |
| HD 122064 | K3 | 0.80 | 43 |
| HD 122120 | K5 | 0.71 | 36 |
| HD 125455 | K1 | 0.79 | 20 |
| HD 126053 | G1 | 0.86 | 30 |
| HD 127334 | G5 | 1.10 | 24 |
| HD 130992 | K3 | 0.77 | 36 |
| HD 132142 | K1 | 0.77 | 21 |

Table S1—Continued

| Name | Spec. Type | Mass ($M_\odot$) | Num. Obs. |
|---|---|---|---|
| HD 136713 | K2 | 0.84 | 79 |
| HD 139323 | K3 | 0.89 | 91 |
| HD 140538 A | G2 | 1.06 | 58 |
| HD 141004 | G0 | 1.14 | 68 |
| HD 143761 | G0 | 1.00 | 29 |
| HD 144579 | G8 | 0.75 | 30 |
| HD 145675 | K0 | 1.00 | 59 |
| HD 145958 A | G8 | 0.91 | 44 |
| HD 145958 B | K0 | 0.88 | 31 |
| HD 146233 | G2 | 1.02 | 52 |
| HD 146362 B | G1 | 1.07 | 29 |
| HD 148467 | K5 | 0.67 | 22 |
| HD 149806 | K0 | 0.94 | 28 |
| HD 151288 | K5 | 0.59 | 22 |
| HD 151541 | K1 | 0.83 | 29 |
| HD 154088 | G8 | 0.97 | 67 |
| HD 154345 | G8 | 0.88 | 53 |
| HD 154363 | K5 | 0.64 | 25 |
| HD 155712 | K0 | 0.79 | 39 |
| HD 156668 | K2 | 0.77 | 93 |
| HD 156985 | K2 | 0.77 | 34 |
| HD 157214 | G0 | 0.91 | 25 |
| HD 157347 | G5 | 0.99 | 46 |
| HD 158633 | K0 | 0.78 | 20 |
| HD 159062 | G5 | 0.94 | 29 |
| HD 159222 | G5 | 1.04 | 55 |
| HD 161797 | G5 | 1.15 | 22 |
| HD 164922 | K0 | 0.94 | 50 |
| HD 166620 | K2 | 0.76 | 35 |
| HD 168009 | G2 | 1.02 | 24 |
| HD 170493 | K3 | 0.81 | 33 |

Table S1—Continued

| Name | Spec. Type | Mass ($M_\odot$) | Num. Obs. |
|---|---|---|---|
| HD 172051 | G5 | 0.87 | 28 |
| HD 176377 | G0 | 0.92 | 32 |
| HD 179957 | G4 | 1.01 | 39 |
| HD 179958 | G4 | 1.03 | 38 |
| HD 182488 | G8 | 0.96 | 45 |
| HD 182572 | G8 | 1.14 | 27 |
| HD 185144 | K0 | 0.80 | 122 |
| HD 185414 | G0 | 1.07 | 27 |
| HD 186408 | G1 | 1.07 | 35 |
| HD 186427 | G3 | 0.99 | 44 |
| HD 190067 | G7 | 0.80 | 50 |
| HD 190360 | G6 | 1.01 | 45 |
| HD 190404 | K1 | 0.70 | 21 |
| HD 190406 | G1 | 1.09 | 32 |
| HD 191785 | K1 | 0.83 | 22 |
| HD 191408 | K3 | 0.69 | 36 |
| HD 192310 | K0 | 0.82 | 45 |
| HD 193202 | K5 | 0.67 | 38 |
| HD 196761 | G8 | 0.83 | 27 |
| HD 197076 | G5 | 0.99 | 86 |
| HD 201091 | K5 | 0.66 | 64 |
| HD 201092 | K7 | 0.54 | 62 |
| HD 202751 | K2 | 0.75 | 42 |
| HD 204587 | K5 | 0.68 | 20 |
| HD 208313 | K0 | 0.80 | 23 |
| HD 210277 | G0 | 1.01 | 49 |
| HD 210302 | F6 | 1.28 | 23 |
| HD 213042 | K5 | 0.74 | 37 |
| HD 215152 | K0 | 0.78 | 27 |
| HD 216520 | K2 | 0.83 | 60 |
| HD 216259 | K0 | 0.69 | 50 |

Table S1—Continued

| Name | Spec. Type | Mass ($M_\odot$) | Num. Obs. |
|---|---|---|---|
| HD 217014 | G2 | 1.09 | 26 |
| HD 217107 | G8 | 1.10 | 41 |
| HD 218868 | K0 | 0.99 | 53 |
| HD 219134 | K3 | 0.78 | 74 |
| HD 219538 | K2 | 0.81 | 30 |
| HD 219834 B | K2 | 0.82 | 24 |
| HD 220339 | K2 | 0.73 | 36 |
| HD 221354 | K2 | 0.85 | 79 |
| HIP 18280 | K7 | 0.59 | 22 |
| HIP 19165 | K4 | 0.70 | 21 |
| HIP 41689 | K7 | 0.62 | 20 |

Table S2. Detected Planets in the Eta-Earth Survey

| Planet | Star | Period (d) | $M \sin i$ ($M_\oplus$) | Reference |
|---|---|---|---|---|
| 14 Her b | HD 145675 | 1754 | 1651 | (15) |
| 16 Cyg b | HD 186427 | 798 | 521 | (15) |
| 47 UMa b | HD 95128 | 1090 | 826 | (16) |
| 47 UMa c | HD 95128 | 2590 | 252 | (16) |
| 51 Peg b | HD 217014 | 4.2 | 147 | (17) |
| 55 Cnc b | HD 75732 | 14.7 | 264 | (18) |
| 55 Cnc c | HD 75732 | 44.4 | 53.4 | (18) |
| 55 Cnc d | HD 75732 | 5371 | 1241 | (18) |
| 55 Cnc e | HD 75732 | 2.8 | 7.6 | (18) |
| 55 Cnc f | HD 75732 | 261 | 46.3 | (18) |
| 61 Vir b | HD 115617 | 4.2 | 5.1 | (19) |
| 61 Vir c | HD 115617 | 38.0 | 11 | (19) |
| 61 Vir d | HD 115617 | 123 | 23 | (19) |
| 70 Vir b | HD 117176 | 116 | 2372 | (20) |
| HD 1461 b | HD 1461 | 5.8 | 8 | (21) |
| HD 3651 b | HD 3651 | 62.2 | 72.8 | (22) |
| HD 7924 b | HD 7924 | 5.5 | 9.3 | (6) |
| HD 69830 b | HD 69830 | 8.7 | 10.2 | (23) |
| HD 69830 c | HD 69830 | 31.6 | 11.9 | (23) |
| HD 69830 d | HD 69830 | 197 | 17.9 | (23) |
| HD 87883 b | HD 87883 | 2754 | 558 | (24) |
| HD 90156 b | HD 90156 | 49.6 | 16.7 | (25) |
| HD 99492 b | HD 99492 | 17.0 | 33.7 | (15) |
| HD 114783 b | HD 114783 | 493 | 351 | (26) |
| HD 154345 b | HD 154345 | 3341 | 304 | (27) |
| HD 156668 b | HD 156668 | 4.6 | 4.1 | (28) |
| HD 164922 b | HD 164922 | 1155 | 114 | (15) |
| HD 190360 b | HD 190360 | 2915 | 497 | (29) |
| HD 190360 c | HD 190360 | 17.1 | 18.7 | (29) |
| HD 210277 b | HD 210277 | 442 | 405 | (15) |
| HD 217107 b | HD 217107 | 7.1 | 443 | (30) |

Table S2—Continued

| Planet | Star | Period (d) | $M \sin i$ ($M_\oplus$) | Reference |
|---|---|---|---|---|
| HD 217107 c | HD 217107 | 4270 | 831 | (30) |
| ρ CrB b | HD 143761 | 39.8 | 338 | (15) |

Table S3. Candidate Planets in the Eta-Earth Survey

| Name | Period (d) | $M \sin i$ ($M_\oplus$) |
|---|---|---|
| Candidate 1 | 18.1 | 6.5 |
| Candidate 2 | 15.3 | 7.3 |
| Candidate 3 | 9.5 | 8.1 |
| Candidate 4 | 13.7 | 8.9 |
| Candidate 5 | 15.0 | 12 |
| Candidate 6 | 67 | 15 |
| Candidate 7 | 74 | 20 |
| Candidate 8 | 71 | 33 |
| Candidate 9 | 777 | 35 |
| Candidate 10 | 338 | 46 |
| Candidate 11 | 1219 | 62 |
| Candidate 12 | 1697 | 94 |